\documentclass[aps,prb,preprint,showpacs,amsmath,amssymb,groupedaddress]{revtex4-1}

\usepackage{times,graphicx}
\usepackage{dcolumn}
\usepackage{pstricks,pst-node}
\usepackage{mathrsfs}
\usepackage{afterpage}

\usepackage[section]{placeins}
\begin{document}

\title{Two-particle photoemission from strongly correlated systems:   {A dynamical-mean field}  approach}

\author{B.D. Napitu$^{1,2}$,  J. Berakdar$^2$}
\address{$^1$ Max-Planck-Institut f\"ur Mikrostrukturphysik, Weinberg 2, 06120 Halle, Germany\\
$^2$ Institut f\"ur Physik, Martin-Luther-Universit\"at Halle-Wittenberg, 06099 Halle, Germany}

\date{\today}

\begin{abstract}
We study theoretically the simultaneous, photo-induced two-particle excitations
  {of} strongly correlated systems   {on the basis of }the Hubbard model. Under certain
conditions specified in this work, the corresponding transition
probability is related to the two-particle spectral function which we calculate using
three different methods:  the dynamical-mean field theory combined with quantum Monte Carlo (DMFT-QMC) technique, the first order
perturbation theory and the ladder approximations. The results are analyzed and compared
for systems at the verge of the
metal-insulator transitions. The dependencies on the
electronic correlation strength and on doping are explored. In
addition, the account for the orbital degeneracy allows an insight into  the
influence  of  interband correlations on the two particle excitations.   {A} suitable experimental realization
is discussed.
\end{abstract}

\maketitle

\section{\label{sec:level1}Introduction}
Correlation among electrons are at the heart of numerous  phenomena in condensed matter such as
the metal to insulator transition, the emergence of magnetic and orbital ordering and
 high-temperature superconductivity \cite{fulde,march,gebhard}.
Much of today's understanding of the role of electronic correlation is based
on the analysis of the \emph{single} particle quantities, e.g.~the  spectral functions,
and  how these compare  with experimental data \cite{huf}. Among others, a  wide spread experimental technique
for this purpose is the angle-resolved (single) photoemission
spectroscopy, ARPES  \cite{huf}.
On the other hand, two-particle quantities are essential for the study
of important phenomena  such   {as} the optical conductivity \cite{mahan}.
Two particle properties may be classified
in general into  those associated with the particle-hole,   {the} hole-hole and {the} particle-particle
channels;  different techniques are appropriate to access  each of these channels.
Probably the most studied one of them is the
 particle-hole channel \cite{marini} that governs  a number of material
 properties such as  the dielectric and the  optical response \cite{mahan}.
The particle-particle and   {the} hole-hole  channels
have been much discussed in connection with the  Auger electron spectroscopy (AES)
\cite{aesbook,exp1,exp2,exp3,exp4,cini,sawz1,gunnar,marini2,drch,dkud,nolt90} and the appearance potential
spectroscopy (APS) \cite{aps,nolt90}.
Early experimental
works  were focused on simple   {compounds} where the two particle
spectra are well modeled by a convolution the single particle spectra.
 For AES/APS from
 correlated systems several theoretical works \cite{cini,sawz1,gunnar,marini2,drch,nolt90,gonis,seib1,seib2}
 have been put forward for the evaluation of the
two-particle  spectral functions, mostly based on the  Hubbard model \cite{h1,h2,h3}.
 Cini and Sawatzky \cite{cini,sawz1} obtained in their pioneering works
 exact results  for a
completely filled band within the single-band Hubbard model.
 A number of subsequent studies  for arbitrary fillings were conducted, mainly using the
 equation of motion  method and the ladder approximation.
E.g., in the work of Drchal,  the equation of motion   {method} was employed to calculate the
spectral density of   {the} two-particle valence
bands \cite{drch} based on  an approximate   single-particle spectral function.
Other works   \cite{dkud,tdds,nolt90}  utilize the
ladder approximation but differ in their treatments of the single particle quantities.
In the work of Treglia \emph{et al.} \cite{tdds}, the one-particle spectrum is calculated by
evaluating the second order perturbation with respect to the Coulomb interaction and with an
additional local approximation
in order to simplify the calculation.
Drchal and Kudrnovsky  \cite{dkud,nolt90}  employed the self consistent T-Matrix approximation which is valid
at a low electron-occupancy.
Seibold \emph{et al. } \cite{seib1} proposed a
new approach based on the time-dependent Gutzwiller approximation (TDGA) \cite{seib2}
to calculate the electron-pairing;
  they compared also their results with those of the bare ladder approximation (BLA).
\begin{figure}[htb]
\begin{pspicture} (-5,5)
\centering\includegraphics[width=2.8in]{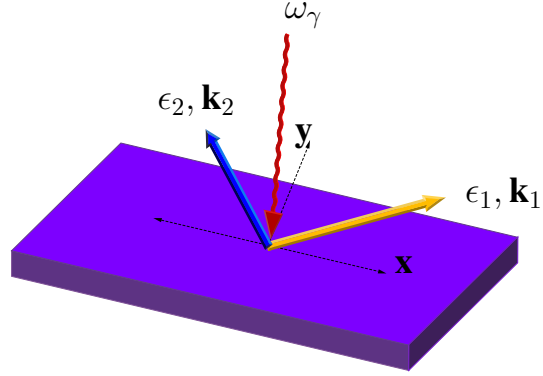}
\end{pspicture}
\caption{A schematics of the one-photon, two-electron  ($\gamma$,2e) experiment. Upon the
absorbtion of a VUV  photon
   with an energy $\omega_\gamma$ two  electrons  are excited into the vacuum and
    simultaneously  detected at the energies $\epsilon_1,\epsilon_2$ and the momenta
       {k}$_1$,  {k}$_2$.}
\label{gbr1}
\end{figure}
These works are mostly discussed in connection with AES and/or APS.
 Recently, an experimental two-particle technique has been developed
 in which two (indistinguishable) valence-band electrons are emitted and detected with
 well defined momenta $\mathbf k_1$ and  $\mathbf k_2$   and specified energies
 $\epsilon_1$ and  $\epsilon_2$ upon the absorption of one single (vacuum ultra violet, VUV) photon \cite{g2e-1,g2e0,g2e1,g2e2}
 (the method is abbreviated by
 $(\gamma, 2e)$, i.e. one VUV photon in, two electrons out), as schematically shown in Fig.\ref{gbr1}.
Excitations by a single electron or positron have also been realized  and a variety of materials
ranging from wide band gap insulators to metals and ferromagnets
\cite{e2e} have been investigated.
The ($\gamma, 2e$) technique is the extension of ARPES to two-particles; from a conceptional point of view one may then
expect to access with ($\gamma, 2e$) the two-particle spectral properties of the valence band, as indeed shown
below explicitly. A distinctive feature of ($\gamma, 2e$) is its vital dependence on
 electronic correlation \cite{jber98}, i.e. the two electrons cannot be emitted with one single photon
 in absence of electronic correlation. The reason for this is the single particle nature of the light-matter
 interaction in the regime where the experiments are performed.
 Theoretical studies concentrated hitherto on  weakly correlated systems  such as simple metals \cite{njjp1,njjp2,njjp3,njjp4,fb}.
 Consequently the two-particle initial state was modeled by a convolution of two single-particle states with the appropriate
 energies. The latter were obtained from conventional band-structure calculations based on the density
  functional theory within the local density approximation.
  Correlation effects were incorporated in the construction of the interacting two-particle states of the emitted
 photoelectrons in the presence of the  crystal potential. An exception to this approach is the study of
 ($\gamma, 2e$) from conventional superconductor where the BCS theory was employed for the initial state \cite{njjp5}.
 While the theory reproduced fairly well the observed experimental trends, the previous theoretical
 formulation will certainly breaks down when dealing with
 strongly correlated materials such transition metal oxides or rare earth compounds with
 partially filled bands. In particular, features akin to the metal-insulator
 transitions are not captured with previous studies.
 Experiments for such materials are currently in preparations.   {Hence,
it is  timely  to inspect the potential of ($\gamma, 2e$) for the study of   strongly
correlated systems. }

In the present work, we will present a  general theory for the two-particle photocurrent
and inspect the conditions under which the  experiment can access information pertinent
to the particle-particle spectral function
in the presence of  strong  electron correlations.  In particular,
we will inspect the particle-particle excitation in  Mott systems
at the verge of  the metallic-insulating transition. For the
description of the properties we employ the
Hubbard model \cite{h1,h2,h3}
 and   {a} non-perturbative technique,   {namely} the
dynamical mean field theory \cite{metvol89,revmod96} (DMFT) in combination with quantum Monte Carlo technique (QMC) \cite{hirsch}.
 For the calculations of the two particle Green function we  will
adopt three different ways: The first one is by
calculating the single and  the two particle spectral functions in the loop of DMFT-QMC self
consistently. This  ensures the fulfillment of   {the} sum rules.
In the second and the third approaches we  basically follow the   methods mentioned above
 for   {the treatment of } AES and APS, i.e.  we consider the
self convolution (first-order perturbation) and the
ladder approximation (LA). We use however the
single particle spectral function as obtained from DMFT.

The  paper is structured as follows:
In  section II we present a general expression for the
two-particle photocurrent and expose its relation to the two-particle Green function.
In section III
 the problem is formulated within the two-band Hubbard model and
 a discussion is presented on how to disentangle matrix elements information from
the ground state two-particle spectral density.
 In section IV and V we present and analyze  the results for the single and two-band
 Hubbard model and compare the results
 obtained at  various level of approximations. Section VI concludes this work.
\section{Correlated two-particle photoemission}
The ($\gamma, 2e$)  set up is schematically shown in Fig.\ref{gbr1}.
These experiments are conducted in the
regime where the radiation field is well described classically and  the time-dependent
perturbation theory in the light-matter interaction  and the dipole approximations
are well justified (low photon density and low photon frequency $\approx 50$ eV). An essential point
for our study is that the operator $D_N$ for the photon-charge coupling is   {a sum }of
 single particle   {operators}, i.e.  $D_N\propto \sum_{i=1}^N \mathbf
A(\mathbf r_i).\hat{\mathbf p}_i$ (in first quantization) where $\mathbf A$ is
the vector potential and $\hat{\mathbf p}_i$ is the momentum operator of
particle $i$. This   {implies that} $D_N$ cannot induce direct many-particle
processes in the absence of inter-particle correlations that help share
among the particles the energy transferred by the photon  to one particle which then results
in multiparticle excitations. A mathematical elaboration on this point
is given in \cite{jber98} and also confirmed below.   { To switch to second quantization
 we write $\Delta=\sum_{mm^\prime}\left<E_{m^\prime}(N_v-2)\right|\mathbf A. (\hat{\mathbf
  p_1}+\hat{\mathbf p_2})\left|E_m(N_v-2)\right>P_2$.  The two-particle photocurrent
($J$), summed over the non-resolved initial and final states $n$ and $m$ is determined according
to the formula \cite{njjp1,njjp2}}

%
\begin{eqnarray}\label{pp11}
J&=&\frac{\alpha_0}{Z}\sum_{N_v}\sum_{mn}e^{-\beta E_n(N_v)}
\left|\left\langle E_m(N_v-2)|\Delta|E_n(N_v)\right\rangle\right|^2 
\delta\left(E-[E_m(N_v-2)-E_n(N_v)]\right)\nonumber\\
%
%
&=& \frac{\alpha_0}{Z}\sum_{N_v}\sum_{mn, m'm''}e^{-\beta E_n(N_v)}
M^\dagger_{mm'} M_{mm''} \langle E_n(N_v)|P_2^\dagger|E_{m'}(N_v-2)\rangle
\langle E_{m''}(N_v-2)|P_2|E_n(N_v)\rangle\nonumber\\
&&\hspace*{2cm}\delta(E-[E_m(N_v-2)-E_n(N_v)]).
\end{eqnarray}
Here we introduced the short-hand notation $M_{kl}$ for the matrix elements. The photon energy  is denoted by
 $E=\hbar\omega_\gamma$, and $\beta$ is the inverse temperature.
  { Furthermore, $\alpha_0=4\pi^2\alpha/\omega_\gamma$, and $\alpha$}
is the fine structure constant.
$P_2=c_\alpha c_\beta$ stands for the (hole-hole) two-particle operator acting on the
 state with
$N_v$ particles with the energy  $E_n(N_v)$. $Z$ is the partition function.
Under certain conditions specified below (the sudden approximation and for high photoelectron energies),
 the variation of the matrix elements,
when we vary $\omega_\gamma$ as to scan the electronic states of the sample, is  smooth in comparison to the change of the matrix elements of $P_2$.
Furthermore, the diagonal elements of $M_{kl}$ are dominant (see below for a justification), i.e. $M_{kl}\approx M$.
In this situation  Eq.(\ref{pp11}) simplifies to ($\rho $ is the density operator).
\begin{eqnarray}\label{ppjb}
&&
J= \frac{\alpha_0}{Z}\sum_{N_v}\sum_{mn}e^{-\beta E_n(N_v)} | M_{mm}|^2
 \langle E_n(N_v)|P_2^\dagger|E_m(N_v-2)\rangle \langle E_{m}(N_v-2)|P_2|E_n(N_v)\rangle\nonumber\\
 &&\hspace*{3cm} \delta(E-[E_m(N_v-2)-E_n(N_v)])\nonumber\\
&& J =\frac{\alpha_0 M^2}{2\pi Z}\sum_{N_v n}\int dt e^{-\beta E_n(N_v)}
 \langle E_n(N_v)|e^{iHt}P_2^\dagger e^{-iHt} \; P_2(t=0)|E_n(N_v)\rangle  e^{iEt}\nonumber\\
  && J=\frac{\alpha_0 M^2}{2\pi Z} \int dt\;
 \mbox{ tr} \left( \rho P_2^\dagger(t) P_2(t=0) \right) \,  e^{iEt}  \nonumber\\
   && J=
  \frac{\alpha_0 M^2}{2\pi} \int dt
 \ll P_2^\dagger(t) P_2(t=0) \gg \,  e^{iEt} .
\end{eqnarray}
On the other hand, from the  spectral decomposition of the two-particle Greens function
\cite{fetka} one infers  for the two-particle spectral density $P(\omega)$ the relation
\begin{eqnarray}
P(\omega)=\frac{\alpha_0}{Z}\sum_{N_v}\sum_{mn}e^{-\beta E_n(N_v)}|\langle E_m(N_v-2)|P_2|E_n(N_v)\rangle|^2(1-e^{-\beta\nu})\delta(\omega-E_m-E_n).
\label{pp12}\end{eqnarray}
 Comparing this equation with Eq.(\ref{ppjb}) we conclude that under the assumption $M_{kl}\approx M$
the photon-frequency dependence of the two-particle photocurrent is proportional to the two-particle spectral density, i.e.
\begin{equation}\label{pp13}
J(\omega)\propto \frac{e^{\beta\omega}}{e^{\beta\omega}-1}P(\omega)
\end{equation}
We recall  that the two particle spectral function   obeys the
sum rule
\begin{equation}\label{pp14}
\int P(\omega)d \omega=\left<n_{\uparrow}n_{\downarrow}\right>.
\end{equation}
A useful auxiliary quantity is partial double  occupancy (up to
a frequency $\Omega$)
\begin{equation}\label{pp15}
K_p(\Omega)=\int^\Omega d\omega P(\omega).
\end{equation}
%
%
%
%
%
%
%
%
\section{Theoretical model}
The aim here is  to explore the potential of the two-particle photoemission for the study of the two-particle correlations
in matter. To do so we start from the generic model that accounts for electronic correlation effects, namely from the
 doubly degenerate Hubbard Hamiltonian. In standard notation we write \cite{h1,h2,h3}
\begin{equation}\label{pp00}
H=\sum_{ij\alpha\sigma}t_\alpha c_{i\alpha\sigma}^\dagger c_{j\alpha\sigma}´+
U\sum_{i\alpha}n_{i\alpha\uparrow}n_{i\alpha\downarrow}+ U^\prime\sum_{i\sigma\sigma'}
 n_{i1\sigma}n_{i2\sigma'},
\end{equation}
where $t_\alpha$ describes hopping between nearest neighbor sites $i,j$ for the
orbitals $\alpha$$\in$(1,2), $U,U' $ stand for the intra- and inter-orbital
Coulomb   {repulsion}, respectively. The above Hamiltonian does not account for the exchange
interaction, pairing and spin flip processes.
The Hubbard
model even for a single band provides an insight into a number of phenomena driven by electronic
correlations   such as
the metal-insulator transition which  cannot be described usually within a static mean
field theory or within an effective single particle picture such as the Kohn-Sham method within
 the density functional theory.
Within the Hubbard
model and for the case of
 infinite connectivity $d\rightarrow \infty$   the self energy turns local \cite{metvol89,hartmann}.
This fact has lead to the development of a new powerful computational scheme
for the treatment of electronic correlation, namely the dynamical mean field theory (DMFT).
For the practical implementation  of DMFT it is essential to map the many-body
 problem onto a single  impurity Hamiltonian
with an additional self consistency relations \cite{geoli92}.
Some of the possible applications of DMFT have been discussed in Ref.[\onlinecite{revmod96}], e.g.
 the long standing problem of the metal insulator transition in the paramagnetic
phase is described in a unified manner.
From a numerical point of view, solving the impurity Hamiltonian is  a challenging task in the self
consistency of DMFT. For this purpose, quantum monte carlo (QMC) methods are shown to be an
effective approach which we will follow  in the present work  with the aim to
  calculate the single and the two particle
Green's functions.   { We note here that since QMC  provides only the data
  for  imaginary times or equivalently at certain Matsubara frequencies,
  we need to perform  analytical continuation to obtain the zero temperatures dynamical
  quantities. This we do by means of the
  maximum entropy method that we implemented using the Bryan method. A detailed
  discussions on this topic can be found in Ref.[\onlinecite{Jarrguber}].}
\subsection{The matrix elements}
Now we have to discuss the validity range of the approximation (\ref{pp12}) that enabled us to assume  for the matrix elements
$M_{kl}\approx M$.
We consider the experiments  in the configuration shown in Fig.\ref{gbr1}. The photoelectron momenta
$\mathbf k_1$ and $\mathbf k_2$  are chosen to be large such that the escape time is shorter than the
lifetime of the hole states. For the description of the photoemission dynamics
 we concentrate therefore  on the degrees of  freedom of the photo-emitted electrons (which amounts to the sudden approximation).
 The energy conservation laws reads then (cf. Fig.\ref{gbr1})
 \begin{equation}
 \hbar \omega_\gamma -\omega=\varepsilon_1 + \varepsilon_2 
 \label{eq:energy}\end{equation}
 where 
  $\omega$ is the initial (correlated) two-particle energy.
   The single particle energies $\varepsilon_j$   {are measured with
  respect  to the edge of the valence band (or with respect to the Fermi level $\mu$ in the metallic case).}
 The matrix elements, e.g. $M_{mm'}$, reduce in the sudden approximation to two particle transition
 matrix elements $M_{if}$. The high energy final state (with energies $\varepsilon_1, \varepsilon_2$)
 we write as a direct product of two
 Bloch states ($\psi_{\mathbf k}$) characterized by the wave vectors $\mathbf k_1$ and $\mathbf k_2$, i.e.
 \begin{equation}
 \Psi_{\mathbf k_1, \mathbf k_2}(\mathbf r_1,\mathbf r_2) =
 \psi_{\mathbf k_1}(\mathbf r_1)\psi_{\mathbf k_2}(\mathbf r_2).\label{end}\end{equation}
 \subsubsection{Intersite ground state correlation}
  Correlation effects enters in the initial two particle states.  In absence of spin-dependent scattering (as is
 the case here) it is advantageous to couple the spins of the two initial states to singlet (zero total spin) and
 triplet (total spin one) states \cite{prl99}. In the paramagnetic phase and if the two electrons are not localized on the same sites
 (they are  mainly around $\mathbf R_i$ and $\mathbf R_j$ with $i\neq j$)
 the initial state is a statistical mixture of singlet and triplet states. The
 radial part we write then as \cite{note1}
 \begin{eqnarray}
 \Psi_{\omega}(\mathbf r_1,\mathbf r_2)&=&\left[
 \varphi_1(\mathbf r_1-\mathbf R_i)\varphi_2(\mathbf r_2-\mathbf R_j)\pm \varphi_1(\mathbf r_2-\mathbf R_i)\varphi_2(\mathbf r_1-\mathbf R_j)\right]
 \chi(|\mathbf r_2-\mathbf r_1 + \mathbf R_i-\mathbf R_j|)\nonumber\\
 &=&\Psi_{\omega}^{(0)}\chi(|\mathbf r_2-\mathbf r_1 + \mathbf R_i-\mathbf R_j| .
 \label{initial}\end{eqnarray}
The "plus" ("minus" sign) stands for the singlet (triplet) channel.
We note that since the transition operator $D_2$ is symmetric with respect to exchange of particles, there is no need
to anti-symmetrize the final state (\ref{end}).
In Eq.(\ref{initial}) the function $\varphi_1(\mathbf r_1-\mathbf R_i)$ and $\varphi_1(\mathbf r_2-\mathbf R_j)$ are
single particle Wannier orbitals localized at the sites $\mathbf R_i$ and $\mathbf R_j$, respectively. $N_i$ is the number of sites and
$\chi(|\mathbf r_2-\mathbf r_1 + \mathbf R_i-\mathbf R_j|)$ is a (dynamical)
correlation factor which we assumed to be
dependent on the relative distance between the electrons. The part $\Psi_{\omega}^{(0)}$ contains
correlation effects due to exchange only. Due to the localization of the Wannier states around the ionic sites
we expect $\chi(|\mathbf r_2-\mathbf r_1 + \mathbf R_i-\mathbf R_j|)$
 to decay with increasing $r_{1/2}$ (for $i\neq j$).
  Since we are dealing with a
lattice periodic problem we can express the  Wannier functions as the Fourier transform of the Bloch
states, i.e.
$\varphi(\mathbf r-\mathbf R_i)=\frac{1}{N_i}\sum_{\mathbf q}^{1.BZ}\psi_{\mathbf q}(\mathbf r)e^{-i{\mathbf q}\cdot\mathbf R_i}$ ($1.BZ$ stands for the first Brillouin zone).
With this relation and exploiting the orthogonality of the Bloch states we obtain upon  straightforward
calculation the following expression for the matrix element
%
 \begin{eqnarray}
 M_{if}&=&\langle \Psi_f|\mathbf A\cdot(\hat {\mathbf p}_1 +\hat {\mathbf p}_2)|\Psi_i\rangle \nonumber\\
 &\approx&\frac {1} {N_i}
 \left\{
 \sum_{\mathbf q_1\mathbf q_2}^{1.BZ} \exp (  -i{\mathbf q_1}\cdot\mathbf R_i  -i{\mathbf q_2}\cdot\mathbf R_j)
 M^{(1)}_{\mathbf q_1, \mathbf k_1}\, \delta_{\mathbf q_2, \mathbf k_2} \pm 1\leftrightarrow 2 \right\}
 \chi(| \mathbf R_i-\mathbf R_j|)\nonumber\\
 &&+\int d^3 r_1d^3 r_2  \; \Psi_{\mathbf k_1, \mathbf k_2}^*(\mathbf r_1,\mathbf r_2)\Psi_{\omega}^{(0)} \:
 \mathbf A\cdot(\hat {\mathbf p}_1 +\hat {\mathbf p}_2)\chi(|\mathbf r_2-\mathbf r_1 + \mathbf R_i-\mathbf R_j|) .\nonumber\\
  \label{matrix_1}.\end{eqnarray}
In this equation $M^{(1)}_{\mathbf q_1, \mathbf k_1}$ is the matrix element for the conventional single photoemission
from the Bloch state $\psi_{\mathbf q_1}$, i.e. $M^{(1)}_{\mathbf q_1, \mathbf k_1}=
\langle \psi_{\mathbf k_1}|\mathbf A\cdot \hat {\mathbf p}_1 |\psi_{\mathbf q_1}\rangle$.
In deriving  the first term of (\ref{matrix_1}) we assumed
 $\chi(|\mathbf r_2-\mathbf r_1 + \mathbf R_i-\mathbf R_j|)$ to vary smoothly with $r_{1/2}$,
 i.e. $\chi(|\mathbf r_2-\mathbf r_1 + \mathbf R_i-\mathbf R_j|)\approx
\chi(| \mathbf R_i-\mathbf R_j|)$ for $i\neq j$.  For 3D periodic structure
  the first two terms of Eq.(\ref{matrix_1}) vanish (momentum and energy conservation laws cannot be satisfied simultaneously).
  Hence, the transition matrix element is determined by the third term of (\ref{matrix_1}), more  precisely by the
  gradient of  the correlation factor $\chi$. If this gradient is smooth on the scale of the variation
  of $\Psi_{\mathbf k_1, \mathbf k_2}$ and/or $\Psi_{\omega}^{(0)}$ then the matrix element vanishes all together
  since $\Psi_{\mathbf k_1, \mathbf k_2}$ and  $\Psi_{\omega}^{(0)}$ are orthogonal. Explicitly we find in this case
  \begin{eqnarray}
  M_{if}&&\approx
\frac {1} {N_i}\sum_{\mathbf q_1\mathbf q_2}^{1.BZ}\left\{ \exp (  -i{\mathbf q_1}\cdot\mathbf R_i  -i{\mathbf q_2}\cdot\mathbf R_j)
\delta_{\mathbf q_2, \mathbf k_2} \delta_{\mathbf q_1, \mathbf k_1} \pm 1\leftrightarrow 2\right\} \nonumber\\
&&\hspace*{2cm}\left.
 \frac{}{ }\mathbf A\cdot(\hat {\mathbf p}_1 +\hat {\mathbf p}_2)\chi(|\mathbf r_2-\mathbf r_1 + \mathbf R_i-\mathbf R_j|)
 \right|_{\mathbf r_2=0=\mathbf r_1} .
\end{eqnarray}
From this expression we conclude that the matrix elements diminish for decreasing correlation $\chi$,  in fact
for $i\neq j$ this contribution to the pair emission is expected to be marginal due to screening.
 \subsubsection{On site ground state correlation}
The major contribution to the matrix   {elements} is expected to stem from the onsite emission $\mathbf R_i = \mathbf R_j$.
Only the singlet state is allowed in the single band Hubbard model. To obtain the two-particle wave function we assume in line of the Hubbard model
that the two electrons scatter via a contact potential of strength $U$ when they are on the same site. The wave function
reads then
\begin{eqnarray}
 \bar\Psi_{\omega}(\mathbf r_1,\mathbf r_2)&=&\left[
 \varphi_1(\mathbf r_1-\mathbf R_i)\varphi_2(\mathbf r_2-\mathbf R_i) +  \varphi_1(\mathbf r_2-\mathbf R_i)\varphi_2(\mathbf r_1-\mathbf R_i)\right]
\bar\chi(|\mathbf r_2-\mathbf r_1|)\nonumber\\ &=&
 \bar\Psi_{\omega}^{(0)}\bar\chi(|\mathbf r_2-\mathbf r_1|).
 \label{initial_onsite}\end{eqnarray}
$\bar\Psi_{\omega}^{(0)}$ describes the  on-site two electron states that include exchange correlation only.
Using only $\bar\Psi_{\omega}^{(0)}$ yields zero matrix elements as shown above.
 To obtain an expression for the
correlation factor $\bar\chi(|\mathbf r_2-\mathbf r_1|)$ (that tends to 1 for $U\to 0$) we
switch to relative $\mathbf R_-$ and center of mass coordinates $\mathbf R_+$. We find that $\bar\chi(|\mathbf r_2-\mathbf r_1|)$
is determined by the integral (Lippmann-Schwinger) equation ($\chi_0$ is determined by asymptotic conditions)
$\bar\chi(R_-) =\chi_0 + U \int d^3 \mathbf R_-^\prime g^r(\mathbf R_-, \mathbf R_-^\prime)\delta^{(3)}(\mathbf R_-^\prime) \bar\chi(\mathbf R_-^\prime)$,
 where $ g^r$ is the retarded Green's function in the relative coordinate.  For (\ref{initial_onsite}) we  find then the explicit solution
\begin{eqnarray}
 \bar\Psi_{\omega}(\mathbf r_1,\mathbf r_2)=
 \bar\Psi_{\omega}^{(0)} (\mathbf r_1,\mathbf r_2)\left[ 1 +  \bar U g^r(\mathbf r_1-\mathbf r_2,0)\right],\quad \bar U=\frac{U}{1-Ug^r(0,0)}.
 {     }\hspace*{1cm}
 \label{wavefunction_onsite}\end{eqnarray}
The key point inferred from this relation is that the two-particle transition amplitude increases
as $U$ increases ($\bar\Psi_{\omega}^{(0)}$  does
not contribute to the matrix elements) and  it vanishes for $U\to 0$.  { It should be noted here that in general
$\bar U $ is a dynamical quantity, as evident from its definitions.}\\
To summarize this section we can say for fixed momenta $\mathbf k_1, \mathbf{k}_2$  of the photoelectrons and
   {for} a given $U$,
 the frequency dependence  of the two-particle emission, $J(\omega)$, is related to the
frequency dependence of the spectral function $P(\omega)$.
 For a given $\omega$, the matrix elements vary with $U$; they  contribute a $\bar U^2$ dependence
 to $J(\omega)$.
 The additional $U$ dependence of  $J(\omega)$ that stems from the spectral
 function will be inspected below.
%
%
%
%
%
%
%
\subsection{Two-particle Green's function}
For our purpose we utilize the general expression for the two particle propagator
\begin{equation}\label{pp01}
\chi^{(\alpha,\sigma),(\alpha^\prime,\sigma^\prime)}_{pp}(\mathbf{q},i\omega_m)= \int\langle T_\tau
c_{\mathbf{k},\alpha^\prime\sigma^\prime}(\tau)c_{\mathbf{q}-\mathbf{k},\alpha,\sigma}(\tau)
c^\dagger_{\mathbf{q}-\mathbf{p},\alpha,\sigma}(0)c^\dagger_{\mathbf{p},\alpha^\prime\sigma^\prime}(0)\rangle .
\end{equation}
 $\int$ is a short-hand notation for $-\sum_{\mathbf{k},\sigma}\int_0^\beta d\tau e^{i\omega_m\tau}$ and
$ \omega_m=\frac{2m\pi}{\beta}$ is the Matsubara frequency, $T_\tau$ is an ordering operator for $\tau$.
The local version of the above two particle Green's function or the onsite \textit{s}-wave   { electron pair}
which will be directly calculated
in the self consistency loop of DMFT-QMC is
\begin{equation}\label{pp02}
\chi_{pp}(\tau)=\left<T_\tau\Delta^\dagger(\tau)\Delta(0)\right>
\quad\mbox{where}\quad \Delta=c_\uparrow c_\downarrow.\end{equation}

The evaluation of the  two particle propagator may be performed with the aid of the
perturbation expansion using the standard diagrammatic theory by selecting the diagrams
 appropriate  for the physical problem at hand.
For the Hubbard model with the short range
 interaction we utilize the ladder-type diagrams.
For  the single-band Hubbard model (an extension to the multi-orbital case is straightforward),
the two particle propagator  reads
\begin{equation}\label{pp03}
\chi_{pp}(\mathbf{q},i\omega_m)=-\frac{1}{\beta}\sum_{\mathbf{k}i \nu_n}
\mathcal{G}(\mathbf{k},i\nu_n)\mathcal{G}(\mathbf{q}-\mathbf{k},i\omega_m-i\nu_n)\Gamma(\mathbf{k},\mathbf{q},i\omega_m).
\end{equation}
We selected the ladder diagrams and summed to all orders. Since in our model the Coulomb
 interaction is static and independent of the wave vector, the vertex function $\Gamma$ reads
\begin{equation}\label{pp04}
\Gamma(\mathbf{k},\mathbf{q},i\omega_m)=1-\frac{U}{\beta}
\sum_{\mathbf{p}i
  \nu_n'}\mathcal{G}(\mathbf{p},i\nu_n')\mathcal{G}(\mathbf{q}-\mathbf{p},i\omega_m-i\nu_n')
\Gamma(\mathbf{p},\mathbf{q},i\nu_n'),
\end{equation}
 meaning that the right hand side of this relation is independent of $\mathbf k$ \cite{nolt90}.
   Thus we obtain
\begin{equation}\label{pp05}
\chi_{pp}(\mathbf{q},i\omega_m)=\frac{\chi(\mathbf{q},i\omega_m)}{1-U\chi(\mathbf{q},i\omega_m)}
\end{equation}
where
\begin{equation}\label{pp06}
\chi(\mathbf{q},i\omega_m)=-\frac{1}{\beta}\sum_{\mathbf{p},i\nu_n}
\mathcal{G}(\mathbf{p},i\nu_n)\mathcal{G}(\mathbf{q}-\mathbf{p},i\omega_m-i\nu_n)
\end{equation}
is the two particle Green's function expressed in terms of the full single particle Green's function.
Performing standard analytical continuation and evaluating the imaginary part of {the}
two particle Green's function one arrives at the following expression for the two particle
spectral function
\begin{equation}\label{pp07}
P(\omega)=\rm{Im}[\chi_{\textit {pp}}(\mathbf{q},\omega+i\delta)] .
\end{equation}
In order to evaluate the above equation, it is sufficient to calculate the
imaginary part of the two particle propagator $\chi(\omega)$, and analytically
continue it to real frequencies. This yields
\begin{equation}\label{pp08}
\chi_i(\omega)=\mathscr{C}_0\int^{\infty}_{-\infty}d\nu\int^{\infty}_{-\infty}d\epsilon
D(\epsilon)\left[A(\epsilon,\nu)A(-\epsilon,\omega-\nu)   (1-f(\nu)-f(\omega-\nu) \right],
\end{equation}
where $\chi_i(\omega)$ stands  for the imaginary part of $\chi(i\omega)$,
$f(\omega)$ is the Fermi distribution function and
$A(\epsilon,\omega)=-\frac{1}{\pi}$Im$\left[\frac{1}{\omega-\epsilon-\Sigma(\omega)}\right]$ is the
full interacting single particle spectral function.
  {$D(\epsilon)$ is      the free density of states and $\mathscr{C}_0$ is
a constant.}
The real part of the  two particle vertex is obtained
via the Kramers-Kronig relation, which follows from the causality condition.
For the case of the degenerate Hubbard model, it is straightforward to extend the above
formulation where now each Green's function contains the composite orbital spin index,
$\alpha=(\alpha,\sigma)$
\begin{equation}\label{pp09}
\chi^{\alpha,\alpha^\prime}_{pp}(\mathbf{q},i\omega_m)=\frac{\chi^{\alpha,\alpha^\prime}(\mathbf{q},i\omega_m)}
{1-U\chi^{\alpha,\alpha^\prime}(\mathbf{q},i\omega_m)}.
\end{equation}
The two particle propagator
$\chi^{\alpha,\alpha^\prime}(\mathbf{q},i\omega_m)$ reads in this case
\begin{equation}\label{pp10}
\chi^{\alpha,\alpha^\prime}(\mathbf{q},i\omega_m)=-\frac{1}{\beta}\sum_{\mathbf{p},i\nu_n}
\mathcal{G}^\alpha(\mathbf{p},i\nu_n)\mathcal{G}^{\alpha^\prime}(\mathbf{q}-\mathbf{p},i\omega_m-i\nu_n) .
\end{equation}

\section{Single Band Hubbard Model}
\begin{figure}[htb]
\centering\includegraphics[width=5in]{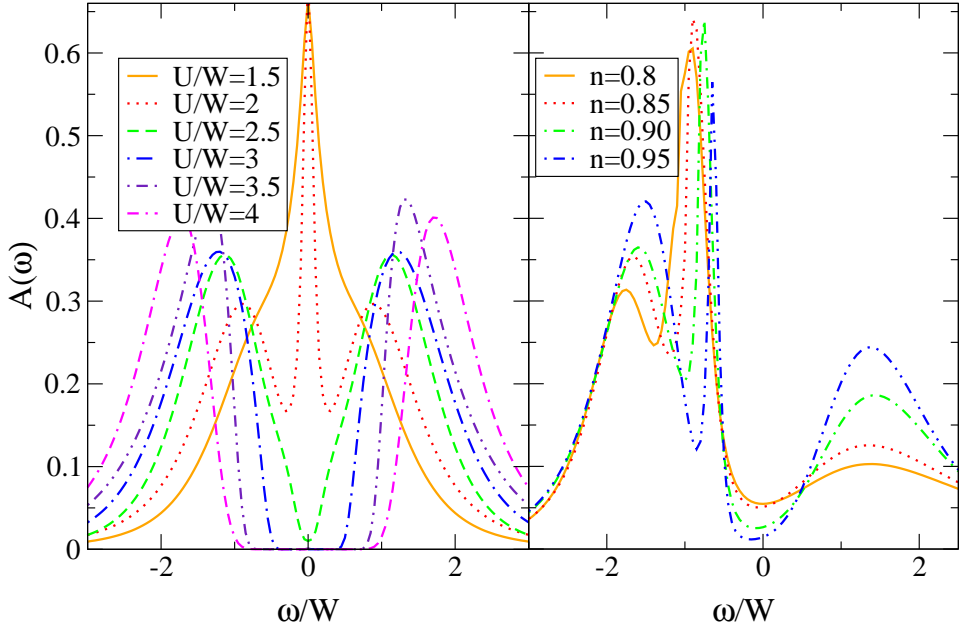}
\psset{unit=0.5}
\caption{The results of the DMFT-QMC for the frequency dependence of the single-particle spectral function $A(\omega)$
for the single-band Hubbard model at
 half filling for various  interaction strengths $U$ (left panel) and for various electron
 occupancy at $U/W=3$ (right panel); for all cases we choose $W=1$.}
\label{res001}
\end{figure}

An essential ingredient for the calculation of the two particle Green's function (\ref{pp08}),
 is the single particle spectral function. The results of the
spectral function obtained from DMFT-QMC method for the
single band Hubbard model at half filling and away from filling are
presented in  Fig.\ref{res001}. In the QMC calculation, the Hirsch-Fye \cite{hirsch}
algorithm is employed with the following  parameters: The energy scale is set by the bandwidth $W=1$.
For the temperature we choose $T/W=0.05$, the increment of time slices
is $\Delta\tau<=$0.5. DMFT-QMC calculations are
performed for the paramagnetic phase and using the Bethe lattice for the free state density in the
self consistency loop.
At half filling   {(the left panel of Fig.\ref{res001})}
the  quasi-particle peak at the Fermi energy is the dominant feature   {in}
 the single particle spectra in the weak coupling interaction signifying
a metallic behavior;  the carriers are itinerant and a
Fermi liquid picture is appropriate.
With an increasing strength of  electronic correlations, localization sets in accompanied by
 a gradual  disappearance of the quasi-particle weight and
the formation of a pseudogap.
Electron transfer between the two bands may occur,  albeit its probability is smaller
than that in the previous case. As the coupling strength further increases,
the gap fully develops indicating an insulating state.

The role of  the double occupancy we inspect by studying the quantity  $\left<n_\uparrow
n_\downarrow\right>$  calculated in the DMFT-QMC
loop.  Evolving from  the weakly interacting (metallic) case  to  the strongly interaction (insulating) phase
 the double occupancy is reduced
\cite{revmod96}, for more
energy is required to overcome the stronger repulsion  whenever forming the double
occupation.
The influence of dopant concentration on the MIT is demonstrated by doping
the insulating phase as depicted in   {the right panel of
  Fig.\ref{res001}}.
  {Contrasting with the results
at  half filling with an interaction strength $U/W=3$, the spectral function
in this case shows a resonance peak at low energies testifying  that
the system  attains again a metallic character. This is because
the doping enhances the number of holes  which in turns increases the
itineracy  such  that the electron can hop  from one site to the other.}

\begin{figure}[t]
\begin{pspicture}(-2.8,9.5)
\centering\includegraphics[width=4.5in, angle=90]{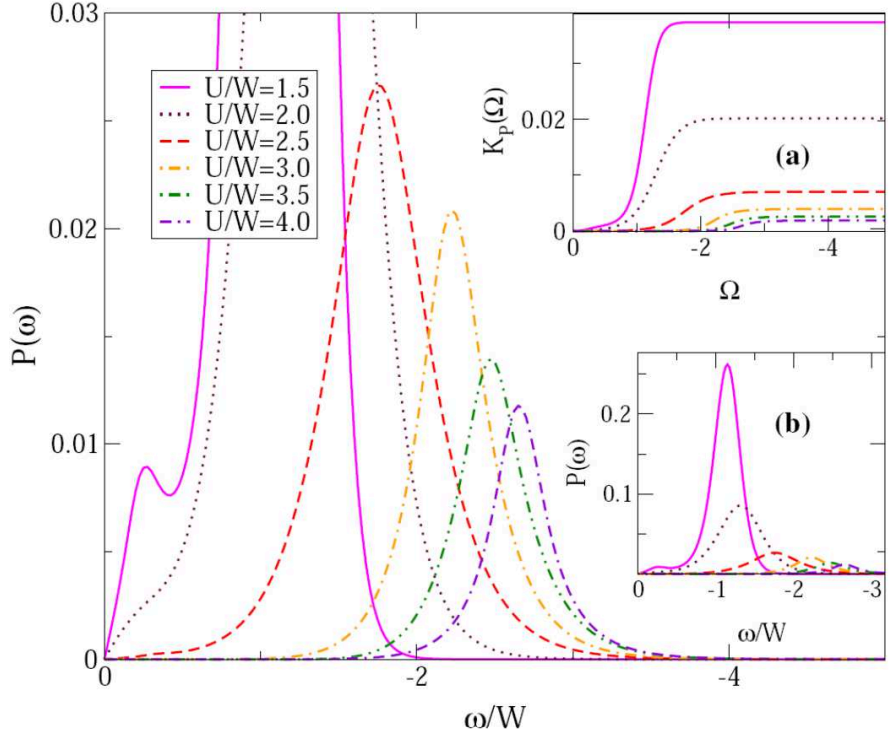}
\end{pspicture}
\caption{Two particle spectral function as function of the correlated, two-particle initial energy $\omega$
 for various interaction strengths. Calculations are performed
  within the self consistency scheme of DMFT-QMC method.
  The large scale figure is shown in inset   {(b)}. Inset   {(a)} shows the integrated spectra using Eq.(\ref{pp15}).
  Note however, that the energy is measured with respect to the uncorrelated
  two-particle Fermi energy $2\mu$.}
\label{res002}
\end{figure}
Having commented on the generic single particle properties of the single band Hubbard
model for Mott systems, we turn now to the discussion of the  particle-particle
spectral function.
  {For small $U/W$ one obtains an intense peak that lies close to $\omega/W=0$.
The origin of such features
can be inferred from the structure of the single particle spectral function: $P$ in this case is
 well modeled by a convolution of two single-particle spectral functions.
A small increase of $U/W$ leads to a reduction of the spectral weight which
shifts the peak to higher energeis (far from $\omega/W=0$). }
As the interaction strength further increases, the spectral weight decreases significantly
signaling  a reduction of double occupation. This argument is
supported by the results of the integrated spectra  depicted in the
inset of Fig.\ref{res002}. In addition to the reduction of the spectral weight,
one also observes the formation of a gap in the low energy regime (near to the zero frequency)
for strong  interaction.  This two particle gap  resembles
the one that appears  in the single particle spectra (cf. Fig.\ref{res001})
which is the usual indicator for an insulating state.
  {We argue here that this is also a signal for the system in the
  insulating state from the point of view of particle-particle excitations.}
\begin{figure}[t]
\centering\includegraphics[width=4.5in]{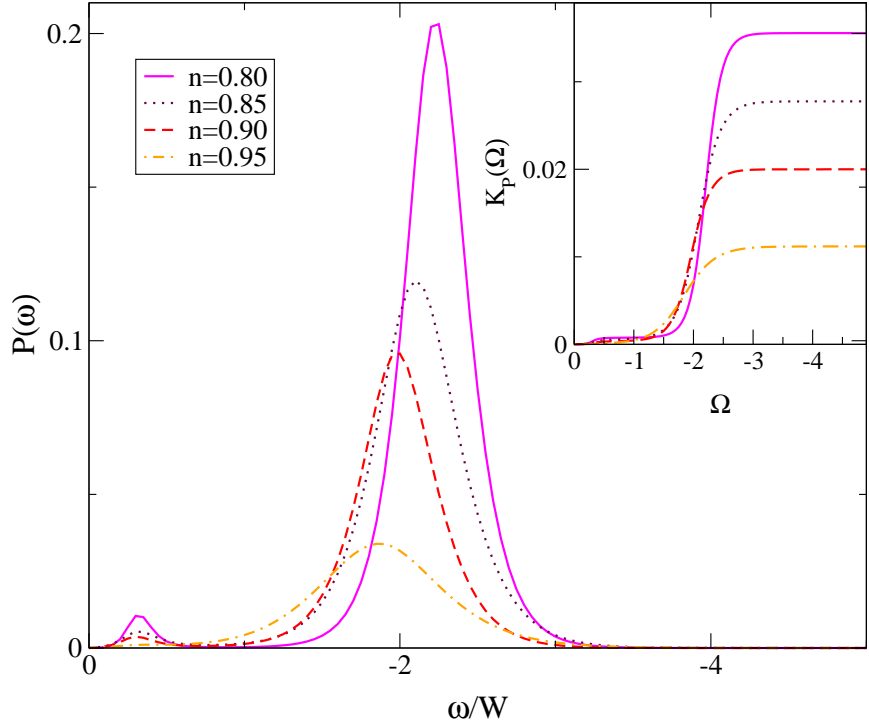}
\caption{ The same as Fig.\ref{res002} with the same notation, however
we inspect here the two particle spectral function away from half filling for
$U/W=3$ and  for   various electron occupancies.
Inset shows the integrated spectra according to equation
(\ref{pp15}).}
\label{res003}
\end{figure}
  {As already pointed out above, the reappearance  of the low energy  resonance
as a function of the doping is a signal for the metallic character and the associated
behavior of the
single particle spectral function.}
The same pattern is also observed in the two particle spectral
function where the strongest peak  occurs in the lowest electron occupancy
and decreases as the Mott insulating phase is approached. Thus the two particle spectra
also highlight the contribution of holes to the double occupancy probability,
and it  is clearly supported by the results for the integrated spectra (see inset of Fig.\ref{res003}).

\subsection{ Relation to the ($\gamma, 2e$) experiments}
 To connect the results of Fig.\ref{res002} to the $(\gamma, 2e)$ signal it is decisive to recall
the statements of equations (\ref{eq:energy},\ref{wavefunction_onsite}):
 The correlated two-particle initial energy that appears in Eq.(\ref{eq:energy}) and which is
   scanned in Fig.\ref{res002}, is in the uncorrelated case merely the  sum of two
   single particle energies $\omega_i$ ($\omega_{uncor}=\omega_1+\omega_2$),
   i.e. in the metallic uncorrelated case we expect some spectral weight
   around $\omega=0$ in Fig.\ref{res002}.   {For a finite $U$, i.e. for a
   correlated system one requires more energy to compensate for the  repulsion of
   the Coulomb interaction.}
   This is the reason  for the shift of the two-particle peak in Fig.\ref{res002} with
   increasing  $U$.   { The same can be observed in the single particle spectral
   function where the distance between the two Hubbard band is approximately on the  order
   of $U/W$.}
   The tendency of larger spectral density with decreasing $U$ is not reflected in the
   $(\gamma, 2e)$ signal $J$. In fact, the opposite will occur. The reason for this is that
   according to eqs. (\ref{wavefunction_onsite}, \ref{ppjb}) $J$ is proportional to the product of
   the matrix elements and the spectral function. On the other hand the matrix
   elements decrease with $U$
     {(cf. eqs.(\ref{matrix_1},\ref{wavefunction_onsite}))}, and in fact vanishes for $U\to 0$ counteracting against the
   trend with $U$ of the spectral function $P$ (cf. Fig.\ref{res002}).
   We stress however, that the
   results  shown in Fig.\ref{res002} are still relevant to the $(\gamma, 2e)$ measurements in that,
   for a given $U$, the matrix elements are hardly dependent on $\omega$.


\subsection{Comparison between different approaches to the two-particle spectral functions}
\begin{figure}[htb]
\begin{pspicture} (-1.5,9.5)
\centering\includegraphics[width=0.4\textwidth]{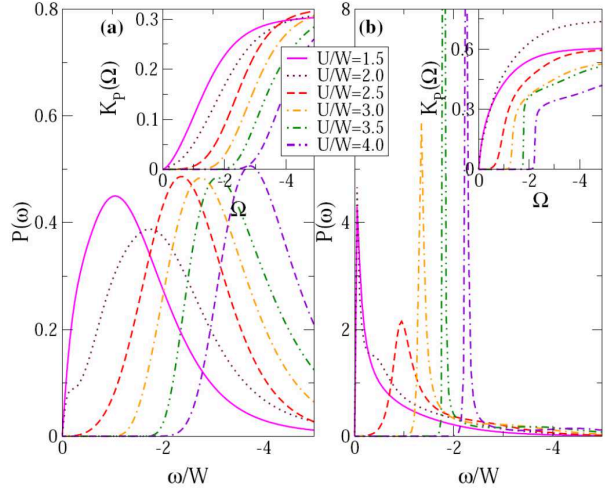}
\end{pspicture}
\caption{The frequency dependence of the two-particle  spectral function at half filling,  calculated with
the first-order perturbation
  approximation   {(a)} and with the full ladder approximation
    {(b)}. Various curves correspond to  different interaction strengths. Inset
  shows the integrated spectra. The single particle quantities are obtained from DMFT.
  Same notation and units as Fig.\ref{res002}.}
\label{res004}
\end{figure}

 To inspect the role of the
 ladder diagram summation (i.e. Eq.(\ref{pp05}) with results in Fig.\ref{res004}  {(b)}),
  we compare with the results (shown in Fig.\ref{res004}  {(a)}) of the
first-order approximation using (\ref{pp08}) (i.e.~with the
convolution of the single particle spectra).
The results of the first order approximation show a smooth, broad
Gaussian-like   feature in the spectra for all
interaction strengths. This is due to the
self-convolution that tends to wash out the  character of the original
function. The presence of a gap in the two-particle spectra highlights
the difference between  the weak and the strong coupling interactions in agreement with
the previous result of DMFT-QMC and with the same energetic origin as discussed above.
That this correct energetic shift is reproduced by  this simple scheme
is the result of using an accurate single particle spectral function.
Another point is the evolution of the two particle spectra from the
weak through the strong coupling limit and the associated behaviour of  the
 spectral weight. In the scheme used in Fig.\ref{res004},
 the weight seems to be comparable  for all
values of the  interaction strengths except for  $U/W=2$ which originates from the low shoulder
in the spectra of Fig.\ref{res001}.
The reduction of the spectral weight is  related
to the probability of the double occupancy. It is then conceivable
to infer that this scheme  violates the sum rule for the two particle spectral
function (which is  dictated by the double occupancy, see Eq.(\ref{pp14}).
This is endorsed by the results for the integrated spectra shown in
the inset of Fig.\ref{res004}  {(a)}. The shift   to higher
frequencies is due to the presence of the gap. No  clear suppression
 is observed as in  Fig.\ref{res002} and Fig.\ref{res003}.

Having obtained the imaginary part of the first order approximation we
 inspect the influence of the ladder diagram  summation on the
two particle spectra. The results are  presented in
Fig.\ref{res004}  {(b)}.
In contrast with previous results obtained in the first order
approximation, the spectra delivered by DMFT-LA are non-uniform
with smooth broad feature and a  satellite
peak.  For the weak interaction strength, the two particle spectra hardly depends on the
 Coulomb interaction strength. As before no
clear  reduction of  the spectral weight is observed.
Interesting features in the DMFT-LA scheme emerge at  higher
interaction strengths, which from
the point of view of the single particle spectra, is already  the regime of
the insulating phase. Instead of suppressing the spectral weight,
the increases of the coupling interaction strength results in  a narrow satellite
peak. The integrated spectra depicted in the inset of Fig.
\ref{res004}  {(b)}
shed some light on  this result.
The integrated spectra within the ladder approximation
exhibit  a suppression
of the weight for higher frequencies in contrast to
 results of the first-order approximation.
We remark that  in the ladder approximation
the suppression of the integrated spectra is not related to a diminishing of
the weight of the two particle spectral function but is associated
 with the width of the spectra that
 become narrow as the interaction increases. All in all we can conclude for these results
 that the $(\gamma, 2e)$ technique is the appropriate tool for testing the validity of
  approximate schemes for the two-particle Green's function.
\begin{figure}[htb]
\begin{pspicture} (-1.5,9.5)
\centering\includegraphics[width=4.5in, angle=90]{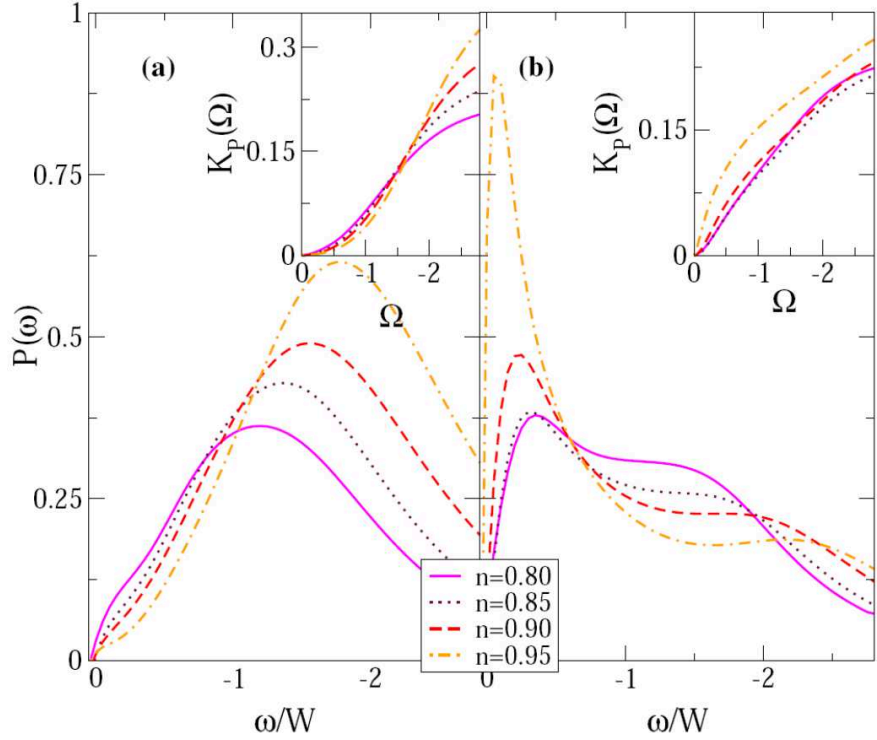}
\end{pspicture}
\caption{ The same as in Fig. (\ref{res004}), however
   we inspect here the role of varying the electron occupancy $n$ at an interaction strength of $U/W=3$.
     {(a)} shows the first-order approximation results whereas in   {(b)} the predictions of the
   ladder approximation are plotted.}
\label{res005}
\end{figure}

The two-particle spectral function away from half filling is
depicted in Fig.\ref{res005} for various occupancies and  for $U/W=3$;
calculations are performed within the first-order approximation and within the
ladder approximation.
No gap formation in the two
particle spectra takes place. This   {is} consistent with  the behaviour of the
single particle spectral function for which
the hole doping of the insulating phase stimulates the formation of quasi-  {particles}.
In the first-order approximation, one obtains the usual broad Gaussian-type structure that
  {diminishes as a function of the} dopant concentration . A somewhat similar
situation is also observed for the results of DMFT-LA.   {In the latter,
however, one observes an intense low-energy peak in the case close to half
filling. The peak decreases as the doping increases. The results of both these approaches arein
contrast to those obtained via  DMFT+QMC where the largest spectral weight is obtained
for the  high doping concentration.
 Therefore, these results do not reflect the fact that
an additional  doping leads to an increase in the double occupancy which
 is clearly supported by the sum
rule results plotted in the inset of Fig.\ref{res005}. Here one observes that the
spectral function at the maximum value of the doping obtains the smallest spectral weight.}
A  similar finding has been  observed in reference \cite{seib1}
where   the bare ladder approximation (BLA) has been utilized.
In their result, the decrease of the electron occupancy
also increases the peaks in the spectra, which they assume to be a violation of the two particle
sum rule.
On the other hand, by using the time-dependent Gutzwiller approximation the opposite
situation occurs: The  two particle spectral weight diminishes as the Mott insulating
phase is approached, which is in line  with what we obtained above within the DMFT-QMC.

\FloatBarrier
\section{Two band isotropic Hubbard model}
The single band Hubbard model   {on which  we based our  above discussion},  is
 useful for systems  with
only a single  band  being close to Fermi energy.
To inspect the role of the  orbital degrees of freedom, which is
known to be important for  the properties of  strongly
correlated systems, a multi-orbital model is needed.  It is the aim of this section
 to study the influence of the
 orbital degeneracy on the single and  two particle spectra.

\begin{figure}[b]
\begin{pspicture} (-2.8,7.5)
\centering\includegraphics[width=4.5in]{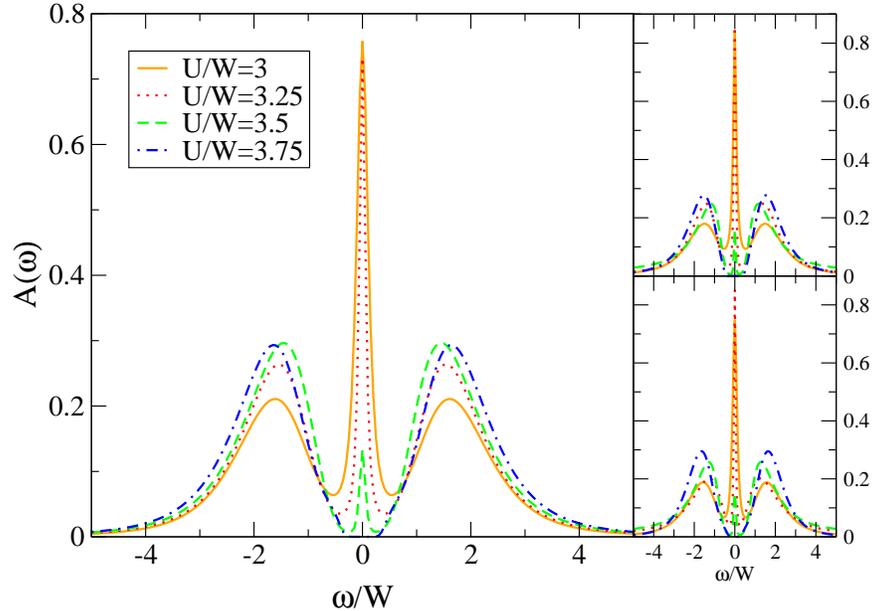}
\end{pspicture}
\caption{
The  DMFT-QMC results  for the frequency dependence of the single-particle spectral function $A(\omega)$
 of the two-band, isotropic Hubbard model  at  half filling. Various curves corresponds to
 different interaction strengths $U$ in units of the band width $W$ (here $W=1$). The insets
show  the orbitally resolved spectral functions for the first (upper inset)
and the second  band (lower inset).}
\label{res006}
\end{figure}

The results for the single particle spectral function within the two band Hubbard model are
presented in Fig.\ref{res006}.
  The results are similar to those obtained within the single band Hubbard model (cf.
Fig.\ref{res002}).
 The metallic phase shows an intense quasi-particle peak
that diminishes as the coupling interaction becomes stronger. The
formation of the gap for a high interaction  strength
 shows the existence of the insulating phase in this degenerate
system. An essential point  that distinguishes the   Mott
transition in the single   {band} from the degenerate band   {case}  is the value of the critical coupling
necessary  to obtain a   dip in the spectral function.
This behavior is well documented  in the works of reference \cite{Lu94}
 employing  the Gutzwiller approximation. There, a relation has been established between the
critical coupling and the orbital degeneracy.
From the orbitally resolved spectral function depicted in the embedded figures,
one also learns that each band undergoes the same transition from metal into
insulating phase. For anisotropic bandwidth
 each band  undergoes an independent metal insulator transition,
a behaviour  coined  as the orbital selective Mott transition  \cite{anis02}.

\begin{figure}[t]
\begin{pspicture} (-1.5,9.5)
\centering\includegraphics[width=4.5in, angle=90]{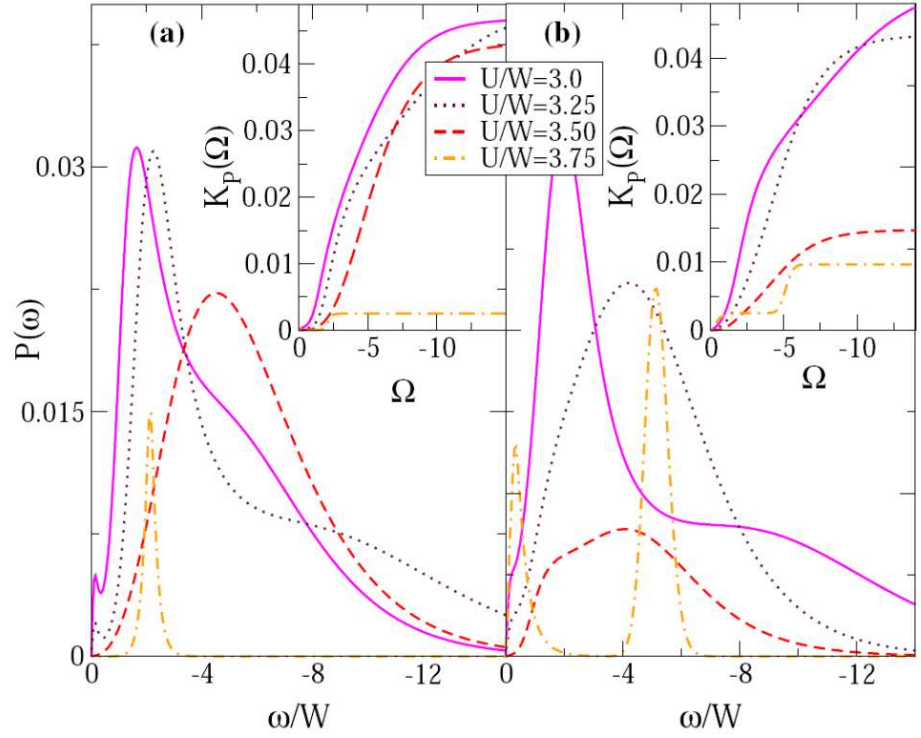}
\end{pspicture}
\caption{
The two-particle spectral function of the degenerate Hubbard model  at half filling
as a function of the two-particle initial energy $\omega$ measured in units of
$W$. The calculations are performed by
the DMFT-QMC method including total bands   {(a)} and inter-band    {(b)}.
The interaction strengths $U$ are varied.}
\label{res007}
\end{figure}

The results of DMFT-QMC   {calculations} for the two particle spectral function are
illustrated  in Fig.\ref{res007} that contains the two spectral functions of the total
band   {(a)} and the interband   {(b)}.
From Fig.\ref{res007} we see that a small increase of the Coulomb
interaction in the weak coupling regime hardly affects
the  overall spectral weight.
  {Furthermore},  increasing  the interaction strength leads however to the reduction of the spectra as
well as to a shift of the  dominant peak to higher energy.

For the case of interband spectra,  there is a clear signal
of the spectral weight reduction already in the metallic case. As
the insulating phase is approached, the two particle spectra show a double-peak
 structure.
\begin{figure}[t]
\begin{pspicture} (-1.5,9.5)
\centering\includegraphics[width=4.5in, angle=90]{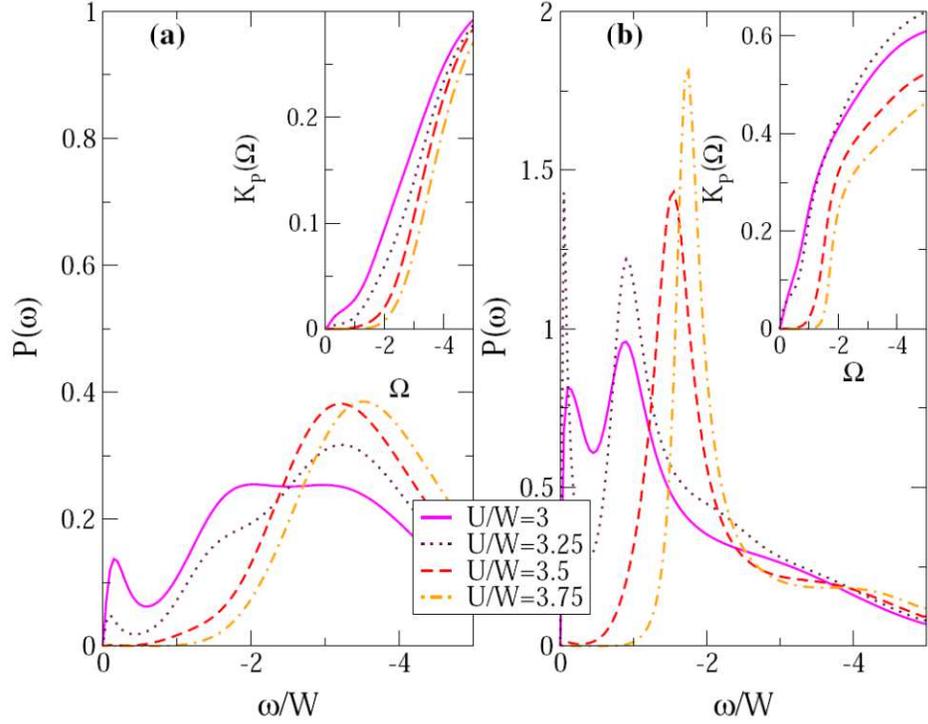}
\end{pspicture}
\caption{
The same as in Fig.(\ref{res007}) for the total bands, however we show in   {(a)} the results
of the   first order perturbation approximation  and in     {(b)}
those of the  ladder approximation.}
\label{res008}
\end{figure}
The two particle spectra obtained by means of the first order approximation as
well as by the ladder approximation are shown in Fig.\ref{res008}  {(a)} and
Fig.\ref{res008}  {(b)} respectively. The behavior of
the two particle spectra in the single band Hubbard model
obtained within the same scheme (see
Fig.\ref{res003}) (e.g. the gap existence, absence of spectral weight reduction)
is also observed in the present case.
In the metallic case however there are new features
predicted  by both approximations namely a  double peak
structure that  disappears in the insulating phase.
Other notable features such as the increase of the weight as the coupling strength increases
 are present in  the results of both methods.
\begin{figure}[htb]
\begin{pspicture} (-1.5,9.5)
\centering\includegraphics[width=4.5in, angle=90]{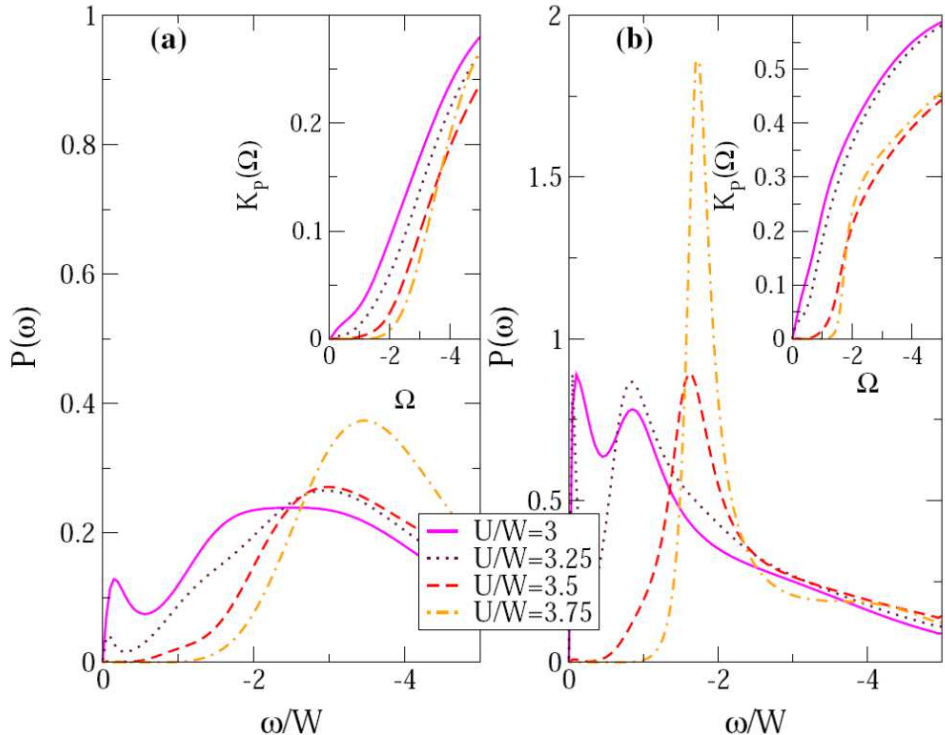}
\end{pspicture}
\caption{The same as in Fig.(\ref{res008}) for the interband  Hubbard model
at  half filling. The figure shows the
results of the  first order approximation   {(a)} and of the ladder approximation
    {(b)}.}
\label{res009}
\end{figure}
The integrated spectra of the degenerate model indicate a violation of   the sum rule for the
 two particle
spectra by both the first order approximation and the
ladder approximation.
From the three scheme: QMC-DMFT, first order and ladder approximations,  the DMFT-QMC methods
provides the more reasonable predictions which practically always obey  the sum rule as a constraint
on the two particle spectral function. This is because,  both the single and the two particle
propagators are calculated on an equal footing in the self consistency DMFT.
An accurate single particle approach when  formulating the  two particle propagator
\cite{nolt90}, does not however
guarantee   the fulfillment of the sum rules. The use of an accurate approach
in the single particle spectra captures however  pertinent features such as the
gap opening in the insulating state which is also observed in the result of DMFT-QMC.

\section{Final remarks and conclusions}
  {We shall now comment on the possible implementation of our proposal. It
  is a widely accepted wisdom that the single band Hubbard model can be employed to
  explain the  results of single particle or particle-hole properties of
  vanadium sesquioxide V$_2$O$_3$. It is shown \cite{revmod96} that the variation of the
  Coulomb repulsion $U$  is realized by
  changing the chemical composition or applying hydrostatic pressure.
  We therefore also argue in this respect that the two-particle properties
  as we have presented here can be accessed in the similar manner.
 For the two band Hubbard model, the results can be again implemented to
 describe the physics of
 V$_2$O$_3$. In this case one can investigate to  role
 of the orbital degrees of freedom.
 The inclusion of orbital degrees of freedom allows the applications of our
 model to wider class of systems.   As for the experimental geometry, our theory is limited
 by the fact that the employed
 self energy is local.   Thus,  our predictions are best  tested by fixing in Fig.\ref{gbr1} the
 momenta of the detected electrons. What should be varied  is then the photon energy $\omega_\gamma$ . Since the momenta
 and the energies of the detected electrons are fixed by the experiment, the only quantity which is
 scanned is the initial correlated two-particle energy $\omega$.  A simulation of the experiments for
 varying momenta and fixed $\omega$ requires an explicitly non-local self-energy which goes beyond the
 validity of the present model.}

To summarize, in this work we explored the potential of two-particle photoemission for the
study of two-particle correlations in correlated
systems. We identified the conditions under which the
two-particle photocurrent is related to the two particle spectral function.
Calculations have been performed  within the framework of the single and the two band Hubbard model.
We performed calculations and compared the  results
of three different schemes DMFT-QMC, the first-order perturbation and
the ladder approximations based on the DMFT single particle spectra.
 In the single band case, the  two particle spectral function
 evaluated with DMFT-QMC is shown to be dependent  on the
  double occupancy in the system.
 As for the single particle spectral function, an increase in the
   electronic correlation strength
results in  suppression of
spectral weight of the
two particle spectra and  in an  opening of a gap near zero two-particle frequency.
The first-order perturbation and  ladder approximation calculations are qualitatively different
from the DMFT-QMC predictions. A finding that can be directly tested by two-particle photoemission spectroscopy.
The inclusion of the orbital degeneracy brings about
an  increase of the critical coupling and
 additional interband contributions to the spectra; these features should also be distinguishable by
  two-particle photoemission experiments.

\section{Acknowledgments}
This work is supported by the international Max-Planck research school for
science and technology of nanostructures and by the DFG under contract SFB
762.\\


\begin{thebibliography}{10}
\bibitem{note1}   {This form of the
 wave function is not the most general one. It neglects three-body (and higher) interactions.
 As discussed in Ref.[\onlinecite{jamal_book}], this doing is well justified for relatively small momentum
 transfer (on the scale of the Fermi momentum), i.e. distant interactions. An exmaple
 for the mathematical expressions
 of the neglected terms is given in \cite{jamN}.}
\end{thebibliography}

\end{document}